\documentclass[10pt, conference]{IEEEtran}
\usepackage[T1]{fontenc}
\usepackage[utf8]{inputenc}

\usepackage{balance}

\ifCLASSINFOpdf
 \usepackage[pdftex]{graphicx}
\else
\fi
\usepackage{url}

\begin{document}
\title{SecureSMART: A Security Architecture\\ for BFT Replication Libraries}


\author{\IEEEauthorblockN{Benedikt H\"ofling, Hans P. Reiser}
\IEEEauthorblockA{Institute of IT-Security and Security Law\\
University of Passau, Germany\\
Email: hoefli06@stud.uni-passau.de, hans.reiser@uni-passau.de}
}


%Telephone: (800) 555--1212, Fax: (888) 555--1212}


\maketitle

\begin{abstract}
Several research projects have shown that Byzantine fault tolerance (BFT)
is practical today in terms of performance.  Deficiencies in other aspects
might still be an obstacle to a more wide-spread deployment in real-world
applications.  One of these aspects is an over-all security architecture
beyond the low-level protocol.  This paper proposes the security architecture \textit{SecureSMART}, which provides dynamic key distribution, internal and
external integrity and confidentiality measures, as well as mechanisms for
availability and access control. For this purpose, it implements
security mechanism among clients, nodes and an external trust center. 
\end{abstract}
%
%% Falls Kürzungsbedarf, evtl. letzten Satz von Abstract weglassen


% \fi
\IEEEpeerreviewmaketitle

\section{Introduction}
Byzantine fault tolerance (BFT) has become more and more practical over the last years.
Besides lot of theoretical BFT research, many research prototypes have made
important contributions. But still, there are only a few complete ready-to-use
prototypes, with \textit{UpRight} \cite{clement09upright} and \textit{BFT-SMART} \cite{sousa12bftsmart} being the most predominant examples. Practical use of
these prototypes in production systems is not common at all.

Clement et al. \cite{clement09making} have criticized the focus of lot of research
systems such as Q/U, HQ and Zyzzyva on optimizing performance in special best-case
situations.  The authors have shown that in common fault situations these systems
easily degrade seriously. They propose a different design of a BFT replication protocol
that guarantees good performance even if there are faulty servers or clients.

Amir et al. \cite{amir08bftattack} describe how timing attacks influence BFT systems. They have extended the two existing criteria \textit{liveness} and \textit{safety} for another criterion \textit{performance-oriented correctness}. They built a new replication protocol that meets this new criterion and handles timing attacks.

Veronese et al. \cite{veronese10ebawa} describe EBAWA, a BFT algorithm that
reduces the communication overhead by using the trusted platform module (TPM) for decreasing the amount of communication steps and replicas. This paper also shows a way to avoid timing attacks of malicious primary nodes by rotating them.

All above-mentioned works show good directions for future BFT developments
beyond the optimization for special best-case situations. However, 
current ready-to-use implementations such as \textit{UpRight} and
\textit{BFT-SMART} do not make use of these results.

In addition, these implementations do not have an elaborate security architecture
for protecting confidentiality, integrity and availability. 
For example, instead of secure mechanisms for dynamic key
management, participant keys typically are defined statically in configuration files.
More advanced mechanisms for access control and Denial of Service (DoS)
protection are desirable.
%So these solutions are not sufficient.
%
In previous work, Amir et al. \cite{amir2005secure} designed \textit{Secure Spread}, which
has an elaborate security concept that shows how to build an architecture that provides
these security mechanism in group communication systems. But it is not aimed for BFT systems.
 
In ongoing research work, we aim at making a contribution to this aspect of BFT
replication.  The basis we use for our work is \textit{BFT-SMART}, an open source Java library
that provides BFT state machine replication \cite{sousa12bftsmart}. The focus of \textit{BFT-SMART} is plainness and robustness. There are also mechanisms that provide authenticity and integrity protection by using preshared keys.
So we are designing and implementing SecureSMART, a security concept for \textit{BFT-SMART} that extends the library by advanced security mechanisms.

\section{Proposed Architecture}

\begin{figure}[tbp]
	\centering
		\includegraphics[width=0.45\textwidth]{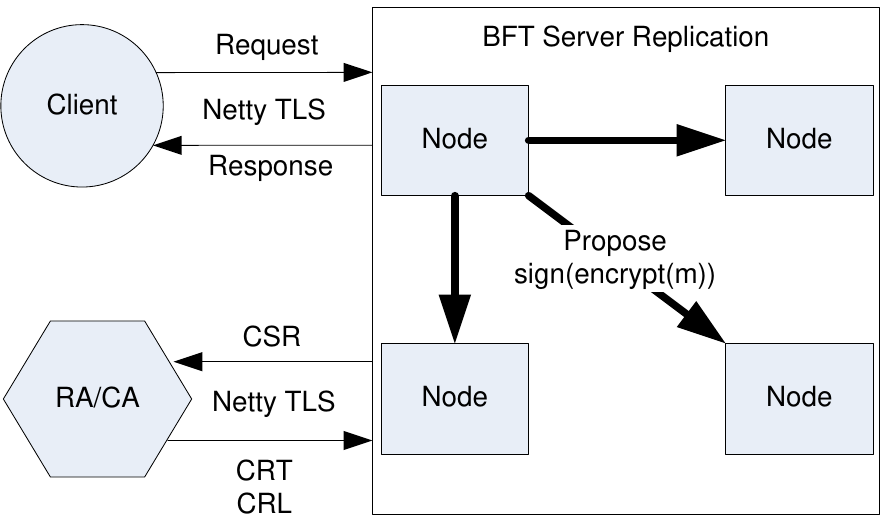}
	\caption{\textit{SecureSMART} architecture}
	\label{fig:architecture}
\end{figure}

Figure \ref{fig:architecture} shows the basic design of our security extension \textit{SecureSMART}.  This architecture meets the following goals:

\begin{itemize}
\item \textbf{Dynamic key distribution:} Whenever a new node wants to join the BFT system it can apply a certificate signing request (CSR) at a registration authority (RA)/certification authority (CA). After receiving the signed certificate the new node can propose its public key to all other nodes. After that a new group key needs to be created, encrypted and proposed to all nodes. In addition, new group keys also need
to be created if a node leaves the BFT system.
\item \textbf{Integrity and confidentiality of internal communication:} The internal communication system of \textit{BFT-SMART} should use the asymmetric key pair to sign and verify every proposed message. The shared group key should be used to encrypt every proposed message which is confidential.
\item \textbf{Integrity and confidentiality of external communication:} The communication system between the clients and the nodes should also be protected. Therefor we can use Transport Layer Security (TLS)
%% Netty hier (ohne Referenz dazu) zu speziell...
% extension from \textit{Netty}
for client-server communication.
\item \textbf{Availability:} 
BFT ensures availability by redundancy, tolerating the failure of a limited number of
$f$ nodes.  But if an adversary can attack the availability of the low-level communication
layer at many or all nodes, the whole system will fail. This means that the BFT system
needs elaborate protection mechanisms against DoS attacks.
\item \textbf{Access control:} There should be a mechanism to blacklist suspicious nodes and clients. One way to do this is a certificate revocation list (CRL). This implies that there is an own certification authority that distributes CRLs to the nodes.  This mechanism also influences the availability.
\end{itemize}

\noindent
We have to address several challenges in order to achieve these
goals:
\balance

\begin{itemize}
\item The dynamic group management brings up some problems: Which nodes may join the BFT system? Is a valid certificate enough for this purpose? How can access control policies be distributed?\\
Another question is how to distribute the public keys? Only the leader node can propose messages including the public key and guarantee that the distributed public key 
will be distributed correctly even in the presence of Byzantine faults.\\
What happens if a node is removed or replaced? The public key of that node is no longer valid for validating messages. This potentially has a big impact on internal validation
procedures within a BFT algorithm.
\item We are using symmetric and asymmetric encryption methods in this security concept which brings the following questions:\\ What are appropriate encryption algorithms? Should the symmetric encryption use block cipher algorithms or stream cipher algorithms? If using a block cipher algorithm, what kind of padding and which mode of operation should be used? 
\item In a BFT scenario each node should receive the same information. Based on this fact, it is advisable to implement group keys for the message encryption. But from this point there raises the question of how the group keys should be generated. Does it make sense if all the nodes together generate the key using Diffie-Hellman (DH)? Is there an advantage by using Tree-based Group Diffie-Hellman (TGDH)? Or is it better if the primary node generates the key?
\item In the case that the group key has leaked there have to be some preventions: How important is perfect forward secrecy (PFS \cite{diffie92pfs}) in this context? When should a new key be established?
\item Measures to improve the protection against DoS attacks are Transmission Control Protocol (TCP) SYN-Cookies and blacklisting of packets source IP addresses. But there might be some problems with the prevention of TCP SYN-Flooding attacks by using SYN-Cookies:\\ At first SYN-Cookies are based deeply in the operating system kernel. Is it possible to enable this option or similar options from the view of the Java Virtual Machine (JVM)?\\ Another problem of using SYN-Cookies might be a performance issue.
\item The RA/CA is a single point of failure. What does this mean in the context of BFT? How can we use BFT replication for the RA/CA?
\end{itemize}

\section{Conclusion}
In terms of performance, BFT systems have already reached a level that
makes them well-suited for real-world applications.  There are still deficits 
in other areas that stop BFT from happening in practice. This work addresses
parts of these problems in the area of security mechanisms.
In this abstract, we have described the basic architecture of \textit{SecureSMART}, 
a security extension for the \textit{BFT-SMART} library we are currently working on,
and we have explained some of the challenges we are handling in this
architecture.


\bibliographystyle{IEEEtran}
\bibliography{IEEEabrv,bib/paper}

\end{document}